\documentclass[submission,copyright]{eptcs}
\providecommand{\event}{TTC 2011} 

\usepackage{graphicx} 
\usepackage[utf8]{inputenc}
\usepackage{url}
\usepackage{amsmath}
\usepackage{amsfonts}
\usepackage{amssymb}
\usepackage{subfigure}
\usepackage{wrapfig}
\usepackage{tabularx}
\usepackage{cite}

\usepackage{color}
\definecolor{listinggray}{rgb}{0.9,0.9,0.9}
\definecolor{keywordcolor}{rgb}{0.5,0,0.1}
\definecolor{commentcolor}{rgb}{0,0.5,0}
\definecolor{stringcolor}{rgb}{0,0,1}
\usepackage[hyper]{listings}

\lstdefinelanguage{viatra}
{morekeywords={@Trigger,trigger,guard,asmfunction,rule,gtrule,if,do,choose,forall,iterate,print,println,log,apply,
    entity,relation,supertypeOf,subtypeOf,typeOf,instanceOf,try,else,pattern,
    precondition,postcondition,action,neg,find,import,namespace,in,below,out,
    inout,let,multiplicity,many_to_one,many_to_many,one_to_many,one_to_one,
    isAggregation,inverse,seq,update,ref,true,false,call,machine,or,
    undef,rename,new,del,delete,move,copy,setValue,setFrom,setTo,with,when,check,change,appear,disappear,upon,cdrule},
 sensitive=true,
 morecomment=[l]{//},
 morecomment=[s]{/*}{*/},
 morestring=[b]{"},
}

\newcommand{\callLink}[1]{\hyperlink{#1}{#1}}
\newcommand{\callAnchor}[1]{\hypertarget{#1}{#1}}

\newcommand{\Viatra}{\textsc{Viatra2}}

\begin{document}
%


\title{Solving the TTC 2011 Reengineering Case with \Viatra{}\thanks{This work
was partially supported by ICT FP7 SecureChange (ICT-FET-231101) European Project.}} \author{{\'A}bel Heged{\"u}s \qquad\qquad Zolt\'an Ujhelyi \qquad\qquad G\'abor Bergmann
\institute{
	Fault Tolerant Systems Research Group \\
	Department of Measurement and Information Systems \\
	Budapest University of Technology and Economics, Hungary \\
	\email{\quad hegedusa@mit.bme.hu \quad\qquad ujhelyiz@mit.bme.hu \quad\qquad bergmann@mit.bme.hu}
	}
}

\def\titlerunning{Solving the TTC 2011 Reengineering Case with \Viatra{}}
\def\authorrunning{{\'A}. Heged{\"u}s, Z. Ujhelyi \& G. Bergmann}

\maketitle

\begin{abstract}
The current paper presents a solution of the \emph{Program Understanding: A Reengineering Case for the Transformation Tool Contest}
 using the \Viatra{} model transformation tool.

\end{abstract}
%

\section{Introduction}
Automated model transformations play an important role in modern model-driven system engineering in order to query, derive and manipulate large, industrial models. Since such transformations are frequently integrated to design environments, they need to provide short reaction time to support software engineers.

The objective of the \Viatra~(VIsual Automated model TRAnsformations~\cite{viatra}) framework is to support the entire life-cycle of model transformations consisting of specification, design, execution, validation and maintenance. 

\emph{Model representation.} \Viatra\ uses the VPM metamodeling approach~\cite{sosym2003_vpm} for describing modeling languages and models. The main reason for selecting VPM instead of a MOF-based metamodeling approach is that VPM supports arbitrary metalevels in the model space. As a direct consequence, models taken from conceptually different domains (and/or technological spaces) can be easily integrated into the VPM model space. The flexibility of VPM is demonstrated by a large number of already existing model importers accepting the models of different BPM formalisms, UML models of various tools, XSD descriptions, and EMF models.

\emph{Graph transformation} (GT) \cite{GT:HandbookII} based tools have been frequently used for specifying and executing complex model transformations. In GT tools, \emph{graph patterns} capture structural conditions and type constraints in a compact visual way. At execution time, these conditions need to be evaluated by \emph{graph pattern matching}, which aims to retrieve one or all matches of a given pattern  to execute a transformation rule. A \emph{graph transformation rule} declaratively specifies a model manipulation operation, that replaces a match of the LHS graph pattern with an image of the RHS pattern.

\emph{Transformation description.} Specification of model transformations in \Viatra\ combines the visual, declarative rule and pattern based paradigm of graph transformation and the very general, high-level formal paradigm of abstract state machines (ASM)~\cite{borger:asm} into a single framework for capturing transformations within and between modeling languages~\cite{scp-2007}. A transformation is defined by an ASM machine that may contain ASM rules (executable command sequences), graph patterns, GT rules, as well as ASM functions for temporary storage. An optional main rule can serve as entry point. For model manipulation and pattern matching, the transformation may rely on the metamodels available in the VPM model space; such references are made easier by namespace imports. 

\emph{Transformation Execution.}
Transformations are executed within the framework  by using the \Viatra\ interpreter. For pattern matching both (i) \emph{local search based pattern matching} (LS) and (ii) \emph{incremental pattern matching} (INC) are available. This feature provides the transformation designer additional opportunities to fine tune the transformation either for faster execution (INC) or lower memory consumption (LS)~\cite{STTT09:AntWorld}.

The rest of the paper is structured as follows. 
 Sec.~\ref{sec:architecture} gives an architectural overview of the transformation, while Sec.~\ref{sec:solution} highlights the interesting parts of our implementation and finally Sec.~\ref{sec:conclusion} concludes the paper.


\section{Solution Architecture}\label{sec:architecture}
\begin{figure}
  \centering
  \includegraphics[width=0.9\textwidth]{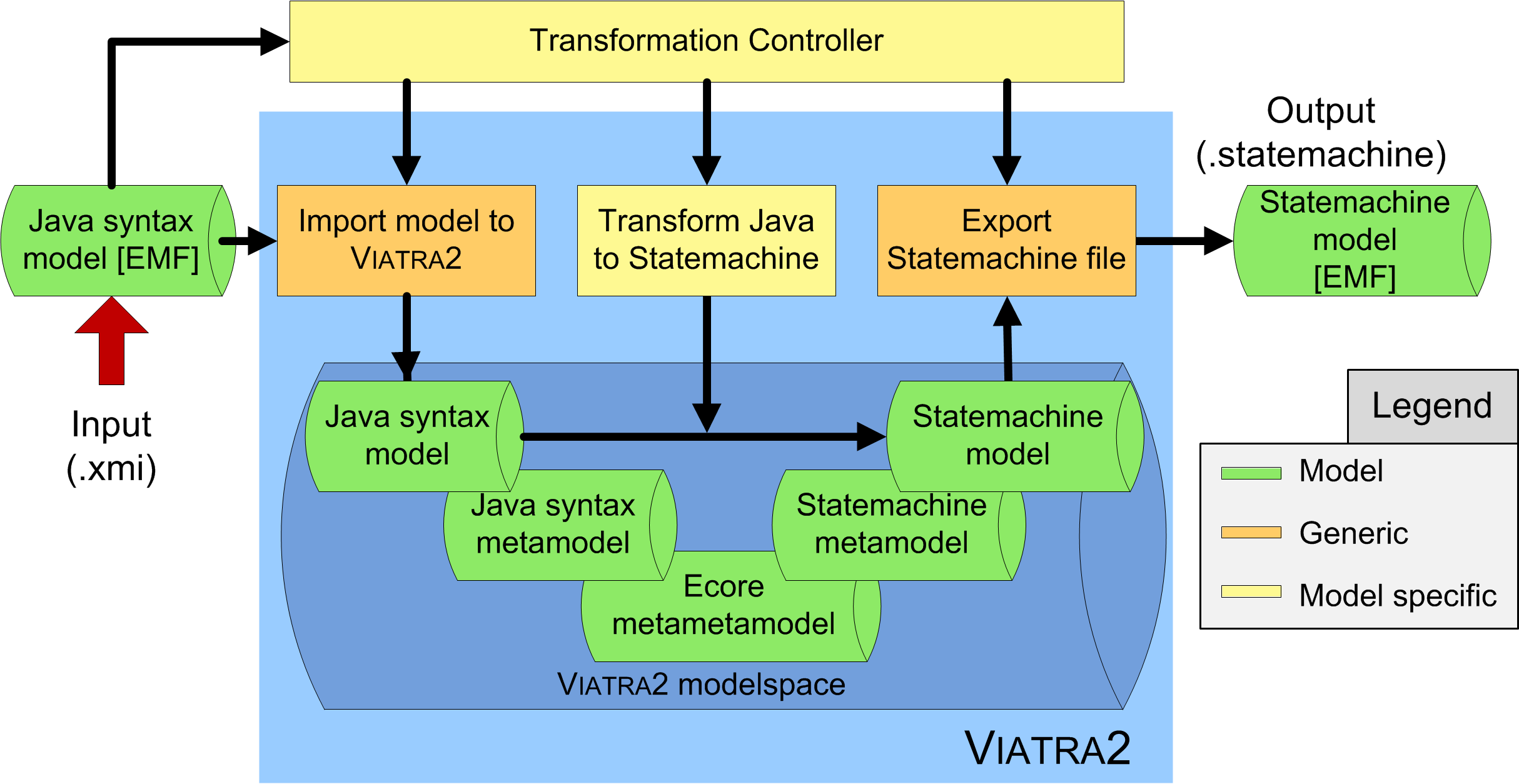}
  \caption{Solution Architecture}\label{fig:arch}
\end{figure}

We implemented our solution for the Program Understanding case study~\cite{programunderstandingcase} using the \Viatra{} model transformation framework. Fig.~\ref{fig:arch} shows the complete architecture with both preexisting (depicted with darker rectangles) and newly created components (lighter rectangles). The optional \emph{Transformation Controller} is an extension to the Eclipse framework that provides an easy-to-use graphical interface for executing the underlying transformation (i.e. it appears as a command in the pop-up menu of XMI files); it is, however, possible to execute the same steps manually on the user interface of \Viatra{}. From the user perspective, the controller is invoked on an \emph{input XMI} file and the result is an \emph{output Statemachine} file.

Note that the transformation is performed on models \emph{inside the VPM modelspace} of \Viatra{} rather than on in-memory EMF models. Although \Viatra{} does not manipulate EMF models directly, it includes a generic support for handling EMF metamodels and instance models.

In order to understand the transformation we briefly outline the metamodeling approach of our solution. The \emph{Ecore metametamodel} is the base of this support, which was defined in accordance with the actual EMF metamodel of Ecore.

Both the \emph{Java syntax graph and Statemachine metamodels} are defined as instances of this metametamodel, and are imported into \Viatra{} with the generic Ecore metamodel importer. Then the input file is used to \emph{import the Java syntax graph into \Viatra{}} and create the \emph{Java syntax model} which is the instance of the Java syntax metamodel.

By executing our implemented transformation, we can \emph{transform the Java syntax model to a Statemachine model} which is an instance of the Statemachine metamodel. This \emph{Statemachine model} is then \emph{exported} to create the output Statemachine file.


\section{Transforming Java syntax to statemachines (J2SM)} \label{sec:solution}


The \emph{J2SM} transformation generates the Statemachine model from the Java syntax graph in the \Viatra{} framework and is implemented in the \Viatra{} Textual Command Language (VTCL)~\cite{sac06_vtcl}. J2SM can be separated into four parts, (1) the construction of the Statemachine states and their outgoing transitions, (2) the processing of triggers and (3) actions for outgoing transitions, and finally (4) connecting the transitions to the target states.

The complete transformation is around $450$ lines of VTCL code including whitespaces and comments (see Appendix \ref{app:xform}). It includes $21$ complex patterns, e.g. the Java class called through an \emph{Instance.activate()} method call can be looked up with the pattern in line~\ref{asm:activateP}. 
%
Finally, the actual manipulation is executed by $5$ declarative rules (e.g. create trigger for a given transition, see line~\ref{asm:trigger}). There are $2$ additional rules for starting and stopping time measurement for different parts of the transformation (see lines~\ref{asm:startTime} and \ref{asm:endTime}).

The transformation starts with a short initialization phase, where the output buffer for the transformation log is cleared, the time measurement starts and a new statemachine model is created.

\paragraph{Construction of states and transitions.} The elements representing the states and transitions of the statemachine are created in the following way:
\begin{enumerate}
  \item First, states are created for each Java class that is not an abstract subclass of the \emph{State} class (see top-level pattern at line~\ref{asm:notAbsState}, called in line~\ref{asm:findState} from a forall construct). A recursive pattern finds these classes by traversing supertype edges. 
  \item Once the state is created, we store the correspondence between the class and the state in an ASM function (essentially a hashmap), the transition handling rule is called (line~\ref{asm:callTrans}).
  \item Since at this point the target states of a transition is probably not available, we only create the \emph{src} and \emph{out} relations.
  \item The transitions in a class are identified by another complex pattern that matches the \emph{Class.Instance.activate()} method calls and finds the called class (see line~\ref{asm:activateP}). The \lstinline{below} keyword is used in a subpattern to express transitive containment of the target class reference within the definition of the source class.  
  \item Once the transition is created, we also store the called class for the transition in the same ASM function to be able to create the \emph{dst} and \emph{in} relations later.
\end{enumerate}

\paragraph{Processing triggers.} Next, the rule handling triggers (see line~\ref{asm:trigger}) is called from line~\ref{asm:callTrigger}. The triggers are created based on the class method, where the \emph{activate()} call is found (see pattern in line~\ref{asm:parMethod}), the switch case constant (line~\ref{asm:parSwConst}) or the catch block exception (line~\ref{asm:parCatch}) that is the closest in the statement hierarchy to the method call. Note that when a catch block is inside another catch block (and similarly for switch cases), the reference solution may choose the outer one for the trigger, while our solution chooses the correct one.

\paragraph{Processing actions.} In the following phase, the action part of the transition is created (line~\ref{asm:action}). The action is created based on the existence of a \emph{send()} method call in the same statement container (found using the pattern in line~\ref{asm:parStmnt}) as the \emph{activate()} call. The name of the action is the same as the enumeration value from the \emph{send()} method call parameter (line~\ref{asm:sendMethodPar}).

\paragraph{Connecting transitions to targets.} Finally, the target of all transitions are handled in the same step using a forall construct (see line~\ref{asm:createTarget}). The interesting part of this rule is the usage of ASM functions to retrieve the correct target state (line~\ref{asm:doubleMagic}). Remember, that the called class is stored for transitions and states are stored for created classes. Therefore, since we iterate through all transitions, the target state can be selected by retrieving the called class for the current transition and the state for that class.

\paragraph{Performance.} We used the provided models to test the performance of our implementation.
 We observed that our framework was unable to handle the biggest model, if we tried to import the complete model, due to \Viatra{}'s VPM representation consuming more memory than EMF. For the other input models, the total runtime of the plug-in loading, import, transformation and export is around 10 seconds, while the transformation itself is around 2 seconds.

However, if we allow a preprocessing phase, which removes unnecessary parts of the model (with the help of EMF IncQuery\footnote{\url{http://viatra.inf.mit.bme.hu/incquery/}}), the big model could be transformed. However, this reduced model is almost equal to the medium model, thus it does not demonstrate the scalability of the approach.

\paragraph{Evaluation.} The transformation handles the core task and both extensions, therefore it is \emph{complete}. The generated state machines are equal to the provided reference solutions, the source and target of transitions are set, while triggers and actions are also created, which means the transformation is \emph{correct}. The transformation code itself is well-structured and is annotated with comments to increase \emph{understandability}. However, it may be challenging for those unfamiliar with the language. Since the language and the framework are not tailored to EMF, the \emph{conciseness} of the transformation is lower and the \emph{performance} of the framework is limited (as discussed above). As a main development direction, we are working on new tools for more powerful EMF support.

%

\section{Conclusion}\label{sec:conclusion}
In the current paper we have presented our \Viatra{} based implementation for the Program Understanding case study~\cite{programunderstandingcase}.


The high points of our transformation are (i) the reusable patterns, (ii) the easily readable transformation language, (iii) the use of ASM functions for easily retrieving corresponding elements, and (iv) that triggers are created for the correct switch case and catch block (as opposed to reference solution).

On the other hand, import-export of models is required and we cannot handle the largest sample input model due to memory constraints.


\bibliographystyle{eptcs}
\bibliography{bib/ttc10,bib/case}

\appendix \newpage
\section{Solution demo and implementation}
Our implementation for the case study together with the current version of \Viatra{} can be installed from the following Eclipse update site: \url{http://mit.bme.hu/~ujhelyiz/viatra/ttc11/}. Additionally, the solution is also available an archive file: \url{http://mit.bme.hu/~ujhelyiz/viatra/ttc11.zip}. Similarly, our solution for the Hello World! case is downloadable from \url{http://mit.bme.hu/~ujhelyiz/viatra/ttc11-helloworld.zip}.

The SHARE image~\cite{share:viatra} usable for demonstration purposes 
 contains our solution for both the Hello World! and Program Understanding cases.

\section{Appendix - Program Understanding transformation}
\label{app:xform}

\lstset{escapeinside={(*}{*)}}
\lstset{numbers=left,stepnumber=10,firstnumber=1}

\begin{lstlisting}[label=lst:trans,caption=Transformation code,numberfirstline=false]
// metamodel imports
import nemf.packages.classifiers;
import nemf.packages.commons;
import nemf.packages.types;
import nemf.packages.modifiers;
import nemf.packages.references;
import nemf.packages.members;
import nemf.packages.statements;
import nemf.packages.parameters;
import nemf.packages.expressions;
import nemf.packages.statemachine;
import nemf.ecore;
import nemf.ecore.datatypes;

@incremental
machine reengineeringJava{(*\label{asm:machine}*)
  
  asmfunction buf/0;    // output buffer
  asmfunction time/1;   // runtime measurement data
  asmfunction models/1; // storing models
  asmfunction sm/1;     // store for statemachine related elements
  
  // entry point of transformation
  rule (*\callAnchor{main}*)() = seq{
    
    // initialize output buffer
    let Buf = clearBuffer("core://reEngineer") in seq{
      update buf() = getBuffer("core://reEngineer");
    }
    
    call (*\callLink{startTimer}*)("main");
    println(buf(), "ReEngineering Transformation started.");
    
    // create new statemachine
    let StateMachine = undef in seq{
      new(StateMachine(StateMachine) in nemf.resources);
      rename(StateMachine,"A_StateMachine");
      update models("sm") = StateMachine;
    }
    
    // finds all State subtypes
    /* 1. A State is a non-abstract Java class (classifiers.Class) that
     extends  the abstract class named ``State'' directly or indirectly.
     All concrete state classes are implemented as singletons [GHJV95]. */
    forall StateClass with find (*\callLink{NotAbstractStateClass}*)(StateClass) do (*\label{asm:findState}*)
     let State = undef, StatesRel = undef, NameRel = undef in seq{
      
      println(buf(), " --> Found State class " + name(StateClass));
      // create states in StateMachine
      new(State(State) in models("sm"));
      new(StateMachine.states(StatesRel,models("sm"),State));
      // store Class -> State correspondence
      update sm(StateClass) = State; 
      // add name to State
      try choose Name with find (*\callLink{NameOfElement}*)(Name,StateClass) do
       let StateName = undef in seq{
        new(EString(StateName) in State);
        setValue(StateName,value(Name));
        rename(State,value(Name));
        new(State.name(NameRel,State,StateName));
      }
      // create transitions from state 
      call (*\callLink{createTransitions}*)(StateClass); (*\label{asm:callTrans}*)
    }
    
    // for each Transition, finds target (use sm map)
    call (*\callLink{createTransitionTargets}*)(); (*\label{asm:callTarget}*)
    
    call (*\callLink{endTimer}*)("main");
    println(buf(), "ReEngineering Transformation ended " + time("main"));
    println(buf(), " RULE: createTransitions ran (in total) for "
     + time("createTransitions"));
    println(buf(), " RULE: createTransitionTargets ran (in total) for "
     + time("createTransitionTargets"));
    println(buf(), " RULE: addTrigger ran (in total) for "
     + time("addTrigger"));
    println(buf(), " RULE: addAction ran (in total) for "
     + time("addAction"));
  }
  
  // finds classes which are subtypes of State
  pattern (*\callAnchor{ClassSubTypeOfState}*)(Class) = {
    Class(Class);
    find (*\callLink{SuperTypeOfClass}*)(SuperType,Class);
    
    find (*\callLink{NameOfElement}*)(Name,SuperType);
    check(value(Name) == "State");
    
  } or { // transitive matching
    Class(Class);
    find (*\callLink{SuperTypeOfClass}*)(SuperType,Class);
    
    find (*\callLink{ClassSubTypeOfState}*)(SuperType);
  }
  
  // restrict subtypes of State to non-abstract ones
  pattern (*\callAnchor{NotAbstractStateClass}*)(Class) = {(*\label{asm:notAbsState}*)
    find (*\callLink{ClassSubTypeOfState}*)(Class);
    neg find (*\callLink{AbstractClass}*)(Class);
  }

  // finds name attribute for element
  pattern (*\callAnchor{NameOfElement}*)(Name,Element) = {
    NamedElement(Element);
    NamedElement.name(NameRel,Element,Name);
    EString(Name);
  }

  // finds supertype of class
  pattern (*\callAnchor{SuperTypeOfClass}*)(SuperType,Class) = {
    Class(Class);
    Class.extends(Extends,Class,NSClassRef);
    find (*\callLink{TargetOfNamespaceClassifierReference}*)(NSClassRef, SuperType);
    Class(SuperType);
  }
  
  // navigate on the classifierReference and target relations to Target
  pattern (*\callAnchor{TargetOfNamespaceClassifierReference}*)(NSClassRef, Target) = {
    NamespaceClassifierReference(NSClassRef);
    NamespaceClassifierReference.classifierReferences(ClassRefRel,
      NSClassRef,ClassRef);
    ClassifierReference(ClassRef);
    ClassifierReference.target(TargetRel,ClassRef,Target);
  }
  
  // matches abstract classes
  pattern (*\callAnchor{AbstractClass}*)(Class) = {
    Class(Class);
    AnnotableAndModifiable.annotationsAndModifiers(ModifierRel,
      Class,Abstract);
    Abstract(Abstract);
  }
  
  // create transitions leading out from StateClass
  rule (*\callAnchor{createTransitions}*)(in StateClass) = seq{(*\label{asm:createTrans}*)
    call (*\callLink{startTimer}*)("createTransitions");
    // finds all transition in class
    /* 2. A Transition is encoded by a methodcall (references.MethodCall),
     which invokes the next state's Instance () method (members.Method)
     returning the singleton instance of that state on which the activate ()
     method is called in turn. This activation may be contained in any of the
     classes' methods with an arbitrary deep nesting. */
    forall ActivateCallClass,ActivateClassRef with
     find (*\callLink{ClassCalledWithActivate}*)(ActivateCallClass,
      ActivateClassRef,StateClass) do let Transition = undef,
       TransRel = undef, SrcRel = undef, OutRel = undef in seq{
      
      println(buf(), "   --> Found activate() methodcall to "
       + name(ActivateCallClass));
      // create Transitions
      new(Transition(Transition) in models("sm"));
      new(StateMachine.transitions(TransRel,models("sm"),Transition));
      
      rename(Transition, name(StateClass) + "-"
       + name(ActivateCallClass));
      // add source, use correspondence for finding state
      new(Transition.src(SrcRel,Transition,sm(StateClass))); 
      new(State.out(OutRel,sm(StateClass),Transition));
      
      // store reference to the class on the other end of transition
      update sm(Transition) = ActivateCallClass; 
      // add trigger
      call (*\callLink{addTrigger}*)(ActivateClassRef, Transition); (*\label{asm:callTrigger}*)
      // add action
      call (*\callLink{addAction}*)(ActivateClassRef, Transition); (*\label{asm:callAction}*)
    }
    call (*\callLink{endTimer}*)("createTransitions");
  }
  
  // finds the class which is called using an activate() method
  pattern (*\callAnchor{ClassCalledWithActivate}*)(ActivateCallClass,
   ActivatedClassRef,StateClass) = { (*\label{asm:activateP}*)
    find (*\callLink{ClassSubTypeOfState}*)(StateClass); // check that the class is a state
    
    // reference to Class
    find (*\callLink{ReferenceTarget}*)(ActivatedClassRef,
     StateClass,ActivateCallClass); 
    Reference.next(ACRNextRef,ActivatedClassRef,InstanceCall);
    // reference to Instance method
    find (*\callLink{MethodCall}*)(InstanceCall,ActivateCallClassInstance); 
    Reference.next(ERNextRef,InstanceCall,ActivateCall);
    find (*\callLink{NameOfElement}*)(Name,ActivateCallClassInstance); // name of Instance
    check(value(Name) == "Instance");
    // reference to activate() method
    find (*\callLink{MethodCall}*)(ActivateCall,ActivateMethod); 
    find (*\callLink{NameOfElement}*)(ActName,ActivateMethod);
    check(value(ActName) == "activate");
  }
  
  // finds reference to target
  pattern (*\callAnchor{ReferenceTarget}*)(TargetRef,SourceElement,ReferencedTarget) = {
    Commentable(SourceElement);
    ReferenceableElement(ReferencedTarget);
    IdentifierReference(TargetRef) below SourceElement;
    ElementReference.target(TargetRefRel,TargetRef,ReferencedTarget);
  }
  
  // finds method called by Caller
  pattern (*\callAnchor{MethodCall}*)(Caller,CalledMethod) = {
    MethodCall(Caller);
    ElementReference.target(TargetRef,Caller,CalledMethod);
    ClassMethod(CalledMethod);
  }
  
  // create references between transitions and target states
  rule (*\callAnchor{createTransitionTargets}*)() = seq{(*\label{asm:createTarget}*)
    call (*\callLink{startTimer}*)("createTransitionTargets");
    println(buf(), " RULE: Creating transition targets");
    
    forall Transition with find (*\callLink{Transition}*)(Transition) do
     let DstRel = undef, InRel = undef in seq{
       println(buf(), " --> Creating target for " + name(Transition));
      // sm(Transition) returns the target class TargetClass
      // sm(TargetClass) returns the corresponding state
      new(Transition.dst(DstRel,Transition,sm(sm(Transition)))); (*\label{asm:doubleMagic}*)
      new(State.in(InRel,sm(sm(Transition)),Transition));
    }
    call (*\callLink{endTimer}*)("createTransitionTargets");
  }
  
  // simple type wrapper for Transition
  pattern (*\callAnchor{Transition}*)(Transition) = {
    Transition(Transition);
  }
  
  // add triggers to transition
  rule (*\callAnchor{addTrigger}*)(in ActivateClassRef,in Transition) = seq{(*\label{asm:trigger}*)
    call (*\callLink{startTimer}*)("addTrigger");
    println(buf(), "  RULE: Creating trigger for " + name(Transition));
    // finds the method where the activate() methodcall happens
    try choose CallingClassMethod with
     find (*\callLink{ParentClassMethod}*)(CallingClassMethod, ActivateClassRef) do
      let Trigger = undef, TriggerRel = undef,
      TriggeringElement = undef in seq{
      println(buf(), "  --> Found class method "
       + name(CallingClassMethod));
      try choose MethodName with
       find (*\callLink{NameOfElement}*)(MethodName,CallingClassMethod) do seq{
    /* 1. If activation of the next state occurs in any method except run(),
     then that method's name (members.Method.name) shall be
     used as the trigger. */
        if(value(MethodName) != "run") seq{
          update TriggeringElement = CallingClassMethod;
        }
    /* 2. If the activation of the next state occurs inside a non-default
     case block (statements.NormalSwitchCase) of a switch statement
     (statements.Switch) in the run() method, then the enumeration con-
     stant (members.EnumConstant) used as condition of the corresponding
     case is the trigger. */
        else seq{
            try choose SwitchCaseConstant with
             find (*\callLink{ParentSwitchCaseConstant}*)(SwitchCaseConstant,
              CallingClassMethod, ActivateClassRef) do
              seq{
                println(buf(), "  --> Found case " + name(SwitchCaseConstant));
                update TriggeringElement = SwitchCaseConstant;
            }
    /* 3. If the activation of the new state occurs inside a catch block
     (statements.CatchBlock) inside the run() method,
       then the trigger is the  name of the caught exception's class.*/
          else try choose CatchBlockClass with
           find (*\callLink{ParentCatchBlockClass}*)(CatchBlockClass,
            CallingClassMethod, ActivateClassRef) do
            seq{
              println(buf(), "  --> Found catch " + name(CatchBlockClass));
              update TriggeringElement = CatchBlockClass;
          }
    /* 4. If none of the three cases above can be matched for the activation
     of the next state, i.e., the activationcall is inside the run() method
     but without a surrounding switch or catch, the corresponding transition
     is triggered unconditionally. In that case, the trigger attribute shall
     be set to --. */
           else seq{
             println(buf(), "  --> Unconditional trigger");
           }
        }
        new(EString(Trigger) in Transition); // creating trigger
        new(Transition.trigger(TriggerRel,Transition,Trigger));
        if(TriggeringElement != undef)
          try choose Name with
           find (*\callLink{NameOfElement}*)(Name,TriggeringElement) do seq{
                // use name of chosen element
                setValue(Trigger,value(Name)); 
          }
        else setValue(Trigger,"--");
      }
     }
     call (*\callLink{endTimer}*)("addTrigger");
    
  }
  
  // finds the class method for a given reference
  pattern (*\callAnchor{ParentClassMethod}*)(CallingClassMethod, IdentifierRef) = {(*\label{asm:parMethod}*)
    ClassMethod(CallingClassMethod);
    IdentifierReference(IdentifierRef) below CallingClassMethod;
  }
  
  // finds the immediate parent switchcase constant for a reference
  pattern (*\callAnchor{ParentSwitchCaseConstant}*)(SwitchCaseConstant,
   ClassMethod, IdentifierRef) = {(*\label{asm:parSwConst}*)
    NormalSwitchCase(NormalSwitchCase);
    // parent switchcase
    find (*\callLink{ParentSwitchCase}*)(NormalSwitchCase,
     ClassMethod, IdentifierRef); 
    // condition of switch
    Conditional.condition(ConditionRel,NormalSwitchCase,Condition); 
    IdentifierReference(Condition);
    EnumConstant(SwitchCaseConstant);
    // referenced constant
    find (*\callLink{ReferenceTarget}*)(Condition,NormalSwitchCase,SwitchCaseConstant); 
  }
  
  // finds immediate parent switchcase, check for lowest parent
  pattern (*\callAnchor{ParentSwitchCase}*)(NormalSwitchCase, ClassMethod, IdentifierRef) = {
    ClassMethod(ClassMethod);
    Switch(Switch) below ClassMethod;
    NormalSwitchCase(NormalSwitchCase);
    Switch.cases(CaseRel,Switch,NormalSwitchCase);
    IdentifierReference(IdentifierRef) below NormalSwitchCase;
    // if there is a lower switch, that must be used
    neg find (*\callLink{LowerSwitch}*)(Switch, IdentifierRef); 
  }
  
  // checks whether a lower switch exists between Switch and the reference
  pattern (*\callAnchor{LowerSwitch}*)(Switch, IdentifierRef) = {
    Switch(Switch);
    Switch(LowerSwitch) below Switch;
    IdentifierReference(IdentifierRef) below LowerSwitch;
  }
  
  // finds the class of the exception used in the parent catch block
  pattern (*\callAnchor{ParentCatchBlockClass}*)(CatchBlockClass, ClassMethod, IdentifierRef) = {(*\label{asm:parCatch}*)
    CatchBlock(CatchBlock);
    // parent catch block
    find (*\callLink{ParentCatchBlock}*)(CatchBlock, ClassMethod, IdentifierRef); 
    
    CatchBlock.parameter(ParRel,CatchBlock,Parameter);
    // targeted parameter
    find (*\callLink{ReferenceTargetOfParameter}*)(Parameter,CatchBlockClass); 
  }
  
  // finds target for parameter through type reference
  pattern (*\callAnchor{ReferenceTargetOfParameter}*)(Parameter,Target) = {
    OrdinaryParameter(Parameter);
    TypedElement.typeReference(TypeRef,Parameter,NSClassRef);
    find (*\callLink{TargetOfNamespaceClassifierReference}*)(NSClassRef, Target);
  }
  
  // finds immediate parent catch block for reference
  pattern (*\callAnchor{ParentCatchBlock}*)(CatchBlock, ClassMethod, IdentifierRef) = {
    ClassMethod(ClassMethod);
    TryBlock(TryBlock) below ClassMethod; // the try block where the catch is
    CatchBlock(CatchBlock);
    TryBlock.catcheBlocks(BlockRef,TryBlock,CatchBlock);
    IdentifierReference(IdentifierRef) below CatchBlock;
    // if there is a lower catch, that must be used
    neg find (*\callLink{LowerCatchBlock}*)(CatchBlock, IdentifierRef); 
  }
  
  // checks whether a lower catch exists between CatchBlock and the reference
  pattern (*\callAnchor{LowerCatchBlock}*)(CatchBlock, IdentifierRef) = {
    CatchBlock(CatchBlock);
    CatchBlock(LowerCatchBlock) below CatchBlock;
    IdentifierReference(IdentifierRef) below LowerCatchBlock;
  }
  
  // add action to transition
  rule (*\callAnchor{addAction}*)(in ActivateClassRef,in Transition) = seq{(*\label{asm:action}*)
    
    call (*\callLink{startTimer}*)("addAction");
    println(buf(), "  RULE: Creating action for " + name(Transition));
    // finds the statement container containing the methodcall
    try choose StatementContainer with
     find (*\callLink{ParentStatementContainer}*)(StatementContainer, ActivateClassRef) do
     let Action = undef, ActionRel = undef in seq{
      println(buf(), "  --> Found container " + name(StatementContainer));
      new(EString(Action) in Transition);
      new(Transition.action(ActionRel,Transition,Action));
    /* 1. If the block (statements.StatementListContainer) containing the ac-
     tivation Call of the next state additionally contains a method Call to the
     send() method, then that call's enumeration constant parameter's name is
       the action. */
      try choose SendMethodParameter with
       find (*\callLink{SendMethodParameterInContainer}*)(SendMethodParameter,
        StatementContainer) do
         try choose Name with
          find (*\callLink{NameOfElement}*)(Name,SendMethodParameter) do seq{
                println(buf(), "  --> Found send () parameter "
                 + name(SendMethodParameter));
                setValue(Action,value(Name));
        }
    /* 2. If there is no Call to send() in the activation call's block,
     the action of the corresponding transition shall be set to --. */
       else seq{
         println(buf(), "      --> No send() in block.");
         setValue(Action,"--");
       }
    } 
    call (*\callLink{endTimer}*)("addAction");
  }
  
  // finds parent statement container
  pattern (*\callAnchor{ParentStatementContainer}*)(StatementContainer, Expression) = {(*\label{asm:parStmnt}*)
    StatementListContainer(StatementContainer);
    ExpressionStatement(Statement);
    StatementListContainer.statements(StatementsRel,
     StatementContainer,Statement);
    ExpressionStatement.expression(ExprRel,Statement,Expression);
    Expression(Expression);
    
  }
  
  /*  finds the EnumConstant used as the Parameter of a send()
   method in a statement container */
  pattern (*\callAnchor{SendMethodParameterInContainer}*)(SendMethodParameter,
   StatementContainer) = {(*\label{asm:sendMethodPar}*)
    StatementListContainer(StatementContainer);
    // parent container
    find (*\callLink{ParentStatementContainer}*)(StatementContainer, SendMethodCall); 
    
    find (*\callLink{MethodCall}*)(SendMethodCall,SendMethod); // methodcall
    find (*\callLink{NameOfElement}*)(SendName,SendMethod);
    check(value(SendName) == "send"); // ensure that it is a send()
    
    find (*\callLink{ArgumentOfMethodCall}*)(Argument,SendMethodCall); // argument of send()
    Reference.next(NextRef,Argument,EnumRef);
    // target of the argument
    find (*\callLink{ReferenceTarget}*)(EnumRef,Argument,SendMethodParameter); 
    
  }
  
  /* finds corresponding arguments for a methodcall */
  pattern (*\callAnchor{ArgumentOfMethodCall}*)(Argument,MethodCall) = {
    MethodCall(MethodCall);
    Argumentable.arguments(ArgRel,MethodCall,Argument);
    Expression(Argument);
  }
  
  /* starts the timer corresponding to the RuleName */
  rule (*\callAnchor{startTimer}*)(in RuleName) = seq{(*\label{asm:startTime}*)
    if(time(RuleName) == undef)
      update time(RuleName) = - systime();
    else
      update time(RuleName) = time(RuleName) - systime();
  }
  
  /* stops the timer corresponding to the RuleName */
  rule (*\callAnchor{endTimer}*)(in RuleName) = seq{(*\label{asm:endTime}*)
    if(time(RuleName) == undef)
      update time(RuleName) = 0;
    else
      update time(RuleName) = time(RuleName) + systime();
  }
  
}
\end{lstlisting}

\end{document}